\documentclass[conference]{IEEEtran}

\usepackage{amsmath,amssymb,amsfonts}
\usepackage{graphicx}
\usepackage{tikz}
 \usepackage{tikz-qtree}
\usetikzlibrary{trees}
\usepackage{booktabs}
\PassOptionsToPackage{hyphens}{url}
\usepackage{hyperref}
\usepackage{url}
\usepackage{filecontents}
\usepackage{cite}
\usepackage{algorithm}
\usepackage{algpseudocode}
\usepackage{listings}
\algnewcommand\algorithmicswitch{\textbf{switch}}
\algnewcommand\algorithmiccase{\textbf{case}}
\algnewcommand\algorithmicassert{\texttt{assert}}
\algnewcommand\Assert[1]{\State \algorithmicassert(#1)}%
\algdef{SE}[SWITCH]{Switch}{EndSwitch}[1]{\algorithmicswitch\ #1\ \algorithmicdo}{\algorithmicend\ \algorithmicswitch}%
\algdef{SE}[CASE]{Case}{EndCase}[1]{\algorithmiccase\ #1}{\algorithmicend\ \algorithmiccase}%
\algtext*{EndSwitch}%
\algtext*{EndCase}%
\usepackage[margin=2cm]{geometry}
\usepackage[most]{tcolorbox}
\newtcolorbox{mytextbox}[1][]{%
  sharp corners,
  enhanced,
  colback=white,
  height=6.5cm,
  attach title to upper,
  #1
}
\usepackage{pgfplots}
\usepackage{tikz}
\newcommand*\circledq[1]{\tikz[baseline=(char.base)]{
           \node[shape=circle,draw,inner sep=0.2pt] (char) {#1};}}
\newcommand*\circled[1]{\tikz[baseline=(char.base)]{%
           \node[shape=circle,fill=green!15,draw,inner sep=0.1pt] (char) {#1};}}

\usepackage{enumitem}
\usepackage{xcolor}
\newtcolorbox{mytextbox2}[1][]{%
  sharp corners,
  enhanced,
  colback=white,
  height=12cm,
  attach title to upper,
  #1
}
\def\BibTeX{{\rm B\kern-.05em{\sc i\kern-.025em b}\kern-.08em
    T\kern-.1667em\lower.7ex\hbox{E}\kern-.125emX}}
\begin{document}
\title{\emph{SRACARE}: Secure Remote Attestation \\  with Code Authentication and Resilience Engine}
\author{\IEEEauthorblockN{Avani Dave, Nilanjan Banerjee and Chintan Patel}
\IEEEauthorblockA{Department of Computer Science and Electrical Engineering, University of Maryland, Baltimore County}
}
\maketitle
\begin{abstract}
\par Recent technological advancements have enabled proliferated use of small embedded and IoT devices for collecting, processing, and transferring the security-critical information and user data. This exponential use has acted as a catalyst in the recent growth of sophisticated attacks such as the replay, man-in-the-middle, and malicious code modification to slink, leak, tweak or exploit the security-critical information in malevolent activities. Therefore, secure communication and software state assurance (at run-time and boot-time) of the device has emerged as open security problems. Furthermore, these devices need to have an appropriate recovery mechanism to bring them back to the known-good operational state. Previous researchers have demonstrated independent methods for attack detection and safeguard. However, the majority of them lack in providing onboard system recovery and secure communication techniques. To bridge this gap, this manuscript proposes \emph{SRACARE} - a framework that utilizes the custom lightweight, secure communication protocol that performs remote/local attestation, and secure boot with an onboard resilience recovery mechanism to protect the devices from the above-mentioned attacks. The prototype employs an efficient lightweight, low-power 32-bit RISC-V processor, secure communication protocol, code authentication, and resilience engine running on the Artix~7 Field Programmable Gate Array (FPGA) board. This work presents the performance evaluation and state-of-the-art comparison results, which shows promising resilience to attacks and demonstrate the novel protection mechanism with onboard recovery. The framework achieves these with only 8$\%$ performance overhead and a very small increase in hardware-software footprint.\\
\end{abstract}
\begin{IEEEkeywords}
Secure-boot, Remote attestation, Embedded System Architecture, IoT devices, Attack resilience, Fault Tolerant and Trusted Embedded Systems, Intelligent Embedded Systems 
\end{IEEEkeywords}
\section{Introduction} \label{Intro}
\par The recent technological advancement has \mbox{catastrophically} increased the utilization of small embedded and IoT devices in applications ranging from industrial control systems, vehicular systems, and home automation systems. Attack \cite{furtak:2014} has demonstrated that the software on these devices can be compromised even when powered off. Remote malware attacks such as Stuxnet \cite{Stuxnet} and Jeep \cite{jeep} can modify the firmware or software of the device. Other attacks such as man-in-the-middle \cite{Man1}, record \& replay \cite{Replay} have shown that the security-critical information of the device can be leaked, modified, and utilized for malevolent activities. Attacks such as Denial of Service (DoS) \cite{DoS} can flood the communication interface of an application to disrupt or damage its normal operation. Therefore, secure communication and software state assurance (at run-time and boot-time) has become paramount essential for the system's security assurance. Unfortunately, the small embedded and IoT systems are computationally weak and do not have in-built security and integrity checking primitives. Hence, they are an active substrate for cyber-attacks that violates software integrity or use the leaked critical information in malicious activities.  
\par Secure boot is a process of measuring the {\bf{boot-time}} integrity and authenticity of the software running on the device. It assures that the device boots up with an untampered and authorized software provided by a legitimate vendor. Thus, a secure boot process becomes a critical step in the device firmware and software security at boot-time. The Remote Attestation (RA) is a popular method of detecting the malicious code presence on the device. RA is a client-server protocol between an untrusted Prover (P\textsubscript{r}) and a remote trusted Verifier (V\textsubscript{r}) devices. The V\textsubscript{r} requests the P\textsubscript{r} for the proof of integrity and/or authenticity of the current state of the device software at run-time, P\textsubscript{r} performs the appropriate checks and sends the report to the V\textsubscript{r}. The V\textsubscript{r} validates the integrity and authenticity of the current state of the software on the P\textsubscript{r}. Previous work has demonstrated RA's utility for software updates \cite{Seshadri2006} and deletion\cite{Perito2010}. The conventional secure boot and RA systems often stop or reset the device upon detecting the malevolent code presence. The device requires the code reflash to restore it to the normal operational state. These brings the requirement of recovery/reflash logic.
\par While conventional RA systems provide run-time detection of corrupted software state, it suffers from the following limitations: 
\begin{itemize}
\item{} It does not provide an efficient recovery mechanism. \item{} It is used for detecting the integrity and authenticity of the system's software state at run-time and not at boot-time. The system needs a secure boot for boot-time software state assurance. \item{} The majority of conventional RA systems did not have the secure communication protocol between the RA devices, which can result in leakage of cryptographic keys, information, or reports and potential misuse by the adversaries. 
\end{itemize}
\par The conventional secure boot and RA devices perform the recovery either by reflashing the application code memory over-the-air or manually. The former method is prone to man-in-the-middle \cite{Man1} and replay \cite{Replay} type of attacks. A smart attacker can corrupt the networking stack to fail the over-the-air code reflash. These necessitates manual intervention for the recovery of the device. Currently used embedded and IoT devices have a wide variety of applications ranging from ships, industrial plants, CCTV cameras, and distributed network sensors (such as dust nodes). These applications require the placement of these devices into the locations that are relatively hard to access for performing device maintenance. Therefore, those devices require an {\bf{onboard recovery mechanism}} to reduce the downtime and maintenance cost. Recent implementations such as \cite{Healed:2019} and \cite{Secerase:2010} attempt to provide recovery methods to the affected devices. However, they lack in providing either the secure boot or RA implementation.
\par To address these limitations, the proposed work presents \emph{SRACARE} a framework that uses a custom communication protocol to measure (run-time and boot-time) integrity and authenticity of the device's software. It also presents a novel resilience and onboard recovery method to protect the device from the malicious code modification attacks.
\begin{figure}[h]
\begin{center}
\includegraphics[width=3.2in]{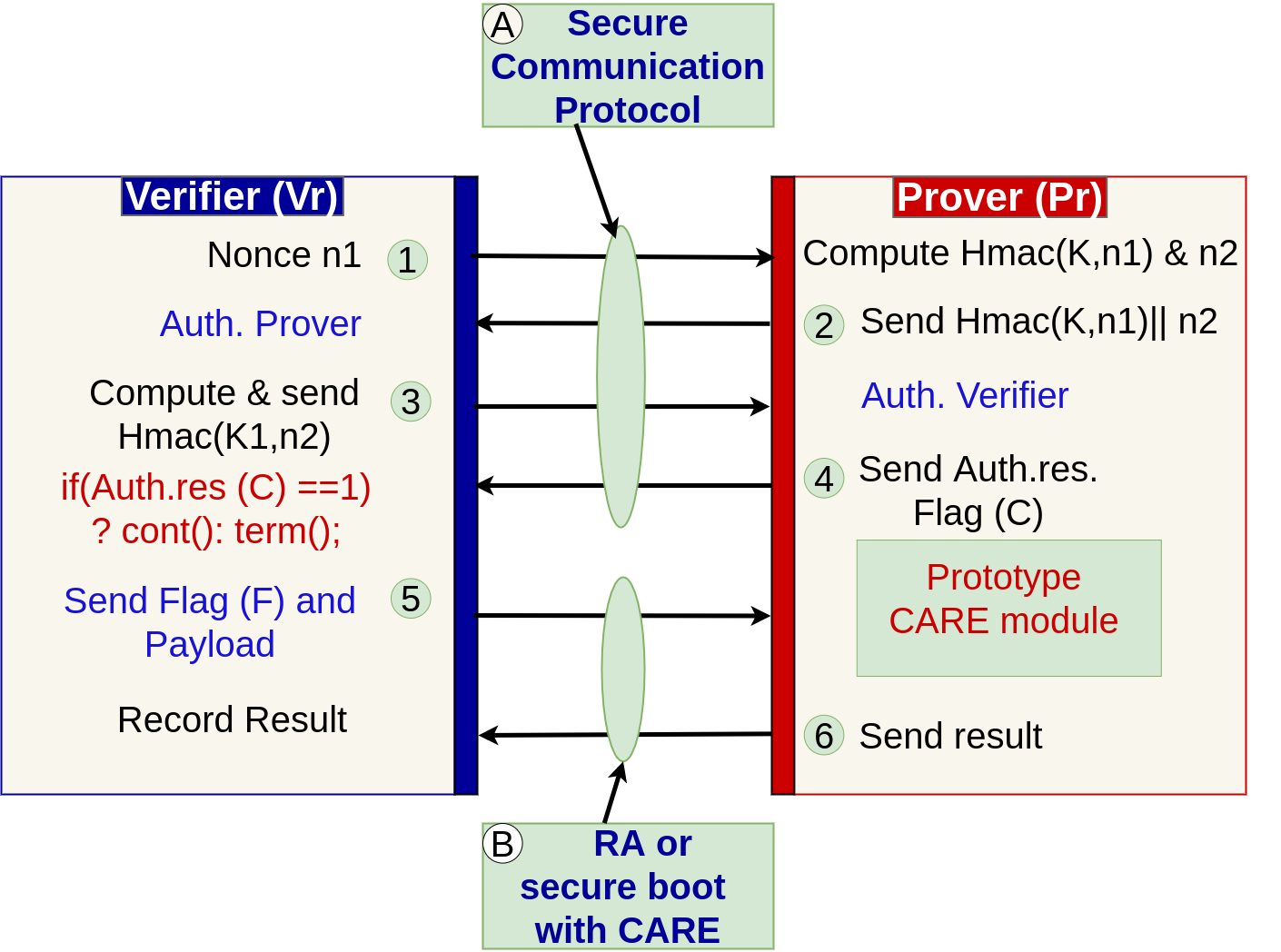}
\end{center}
\vspace{-2ex}
\caption{Highlights the proposed \emph{SRACARE} system design flow and key contributions. It represents the lightweight authenticated secure communication protocol, and a new remote attestation and secure boot architecture using custom \emph{CARE} module.}
\vspace{-2ex}
\label{fig:pro1}
\end{figure} 
\par Fig~\ref{fig:pro1} provides a high-level design overview of the proposed framework. The \emph{SRACARE} system operation is categorized into two sub-tasks: \circledq{A}~Authenticating the trusted V\textsubscript{r} and un-trusted P\textsubscript{r} using lightweight, secure communication protocol and \circledq{B}~Based on the result of task \circledq{A}, it performs either remote (run-time) or secure-boot (boot-time) attestation with onboard recovery by using the prototype \emph{CARE} module. It sends the final computed result to the V\textsubscript{r}, as depicted in step \circled{6} of Fig~1. The authentication of the V\textsubscript{r} and P\textsubscript{r} are performed by sending nonces both ways and computing HMAC (Hash based Message Authentication Code) using a prototyped communication protocol (as shown in steps \circled{1} to \circled{4}). The novelty of this approach resides in the lightweight design of the nonce (n2) generation mechanism at the P\textsubscript{r} side, without requiring the resource heavy nonce generation techniques such as True Random Number Generator (TRNG), details of which are covered in section~\S\ref{SRACAREO} and section~\S\ref{working}. If the authentication check fails the \emph{SRACARE} system sends Flag (C==0) to the V\textsubscript{r} as an attack indicator and V\textsubscript{r} closes the communication between the devices. Upon authentication passing, the P\textsubscript{r} sends Flag (C==1) and the V\textsubscript{r} continues the next step by sending Flag (F) and associated payload to the P\textsubscript{r} device. The P\textsubscript{r} system performs either secure boot with \emph{CARE} or remote attestation based on the value of Flag (F) (as shown in steps \circled{5} and \circled{6}). The (highlighted) prototype \emph{CARE} module provides novel malicious code modification attacks detection, protection and onboard recovery mechanism for small embedded and IoT devices.\\
{\bf{Research Contributions:}} The design and implementation of the proposed \emph{SRACARE} framework presents the following research contributions:
\begin{itemize}
\item {\bf{Secure~Communication~Protocol}:} It demonstrates the prototype implementation of a lightweight secure authenticated communication protocol between the trusted V\textsubscript{r} and untrusted P\textsubscript{r}. 
\item{\bf{Nonce Generation Technique}:} It provides a lightweight novel nonce (n2) generation technique that ensures freshness for the P\textsubscript{r} device without requiring substantial system resources. 
\item{\bf{Remote Attestation~Tools}:} It provides the tools to verify the integrity and authenticity of the state of the software on the device at run-time by the remote V\textsubscript{r}.
\item{\bf{Secure~Boot}:} \emph{SRACARE} presents lightweight, secure boot (boot-time attestation) architecture prototype for \mbox{RISC-V} based small embedded and IoT devices.
\item {\bf{Resilience~Engine}:} It demonstrates the first implementation of the novel resilience engine that provides onboard recovery and protection method to recover the P\textsubscript{r} device after malicious code modification attack.
\item {\bf Prototype~\emph{SRACARE}~Implementation:} It presents the prototype implementation of {\emph SRACARE} framework on FPGA, which can be used as a standalone microcontroller or as a secure boot co-processor such as the Trusted Platform Module (TPM) in large devices.
\end{itemize}

\section{Background and Related Work}
\par This section defines essential concepts used and referenced by the proposed framework, followed by exploring related state-of-the-art works. Measured boot \cite{Arbaugh:1997} is a process of verifying the integrity of the software running on a system. Authenticated boot \cite{UEFI:2019} verifies that the software running on the system is coming from an authorized vendor. The majority of the conventional secure boot and RA techniques perform either measured boot or authenticated boot, and very few support both. The onboard recovery and secure RA communication will be the requirement of the next-generation embedded and IoT devices, as discussed in section~\S\ref{Intro}. Previous researchers have implemented secure boot and/or RA techniques that can be classified into hardware, software, and hybrid approaches:\\
{\bf{Hardware-Based}:} One of the popular methods of secure boot and RA uses a discrete co-processor called the Trusted Platform Module (TPM) \cite{Tpm:2010} recommended by the Trusted Computing Group (TCG). TPM has a special purpose registers called Platform Configuration Registers (PCRs), which cannot be overwritten. However, it can only be extended by hashing the software measurements together with the previous PCR values. The TPM can sign the PCRs with a private attestation key to generate a piece of attestation evidence. The TPM provides hardware root-of-trust. However, it is not suitable for small embedded or IoT devices due to space, size, and cost constraints. Some researchers have used the Trusted Execution Environment (TEE) \cite{Sabt:2015}, Keystone \cite{keystone}, or proprietary implementation of Arm TrustZone \cite{Jiang:2017} for runtime attestation. TrustZone uses two virtual processors called the secure and normal world to enforce the hardware-based isolation. Microsoft's fTPM \cite{mcs:2015} provides a use-case of TrustZone for secure boot and attestation. Intel's SGX \cite{sgx:2013} provides instruction and memory access features that can be used to instantiate protected containers referred to as enclaves by using special instructions and processor extensions. TrustZone and TEE increases the design complexity and cost associated with exclusive licensing. Other secure boot architectures have used complex Unified Extensible Firmware Interface (UEFI) \cite{NSA} design or utilize heavy resources \cite{Haj:2019}, which makes them unsuitable for small embedded and IoT devices.\\
{\bf{Software-Based}:} Researchers have used simulated or software TPM implementations like simTPM \cite{Chakraborty:2019} or IBM's software TPM \cite{Goldman} for secure boot and RA. RISC-V based implementation Sanctum \cite{Sanctum:2018} uses software-based enclaves for attestation. The implementation presented in \cite{ndss:2018} has used cryptographic software core and hash engine for attestation.\\
{\bf{Hybrid}:} SMART \cite{Wong:2018} is a dynamic root-of-trust architecture for low-end devices at runtime. SPM/Sancus: SPM \cite{Stra:2010} and Sancus \cite{sancus:2013} present a security architecture that provides isolation of software modules using additional CPU instructions. TrustLite/TyTAN: TrustLite \cite{TrustLite:2014} and its successor TyTAN \cite{TyTan:2015} provide flexible, hardware-enforced isolation of software modules using Execution-Aware Memory Protection Unit (EA-MPU). Google's latest implementation Opentitan \cite{Ope:2019} provides a secure boot based hardware root of trust. Other RISC-V based secure boot and attestation architecture Shakti-T \cite {Shakti:2017} uses base and bounds concept to secure the pointer's access to the valid memory regions. The existing secure boot and RA architectures are complex \cite{UEFI:2019}, require more resources \cite{Chakraborty:2019}, \cite{Sabt:2015}, \cite{Jiang:2017}, or have been compromised by attacks such as \cite{Vasselle:2017} and \cite{Ang:2017}. 
\par Moreover, none of the available solutions provide the recovery and they use conventional message authentication protocols, which are resource heavy and not suitable for small embedded and IoT devices. Recently implementation Healed \cite{Healed:2019} presents the first recovery mechanism using Merkle Hash Tree (MHT). It assumes that at least one node in the network is untempered, and firmware of that node is used to reflash the corrupted node. Another implementation \cite{Secerase:2010} uses a method of putting software receiver-transmitter code into trusted ROM to connect the device to a recovery server. It requires additional ROM storage, processor, and internal communication bus to be part of Trusted Computing Base (TCB). The proposed \emph{SRACARE} method uses a different approach of storing the recovery data in a secure backup ROM, and it does not require processor and internal communication bus to be part of TCB to reduce the attack surface. 
\vspace{-1em}
\section{Security Model}
\subsection{\bf{Security Properties}} \label{securityprop}
\emph{SRACARE} has derived eight (A1-A8) security properties (from \cite{Vrased:2019}) for the secure boot and remote attestation system design, which are broadly classified into three main domains, namely: Secure Communication, Key Protection, and  Safe Execution.\\
 {\bf{1) Secure Communication}:} The communication protocol between the V\textsubscript{r} and P\textsubscript{r} devices should be resilient to wiretapping and flooding types of attack.\label{seccom}
{\bf{A1.~Eavesdrop Protection}:} The devices should have wiretapping and eavesdropping protection to prevent attacks such as \cite{eaves}, \cite{Replay}, or \cite{Man1} which can cause security-critical information leakage or misuse. 
{\bf{A2.~Flooding Protection}:} It also requires protection from attacks such as \cite{DoS} DoS and \cite{flood} DDoS, which can fail the application.\\
{\bf{2) Key Protection}:} The shared cryptographic secret key (K) and device secrets should not be exposed to the adversary and stored in protected memory with no unauthorized access. {\bf{A3.~Key Confidentiality}:} The secure key (K) must be stored in protected memory or ROM. {\bf{A4.~Access Control Enforcement}:} It requires proper access control policies to prevent unauthorized read-write access to the protected memory.\\
{\bf{3) Safe Execution}:} The design should have an error-free, uninterrupted, and leak-proof implementation and execution of all the system modules. {\bf{A5.~Correct Implementation}:} The implementation of all the submodules should be untempered and correct. {\bf{A6~Atomicity}:} Once triggered, the code execution should not be interrupted. {\bf{A7.~Error Free Execution}:} All the hardware (IPs) and software submodules should have error-free execution. {\bf{A8.~Controlled Invocation}:} The system requires proper code triggering and execution sequence, with no privilege escalation or interruption.
\subsection{\bf{Adversary and Threat Model}}\label{themo}
The adversary: 
\begin{itemize}
\item{}has full control over the firmware and software of the device.
\item{}can perform an unauthorized read or write to the flash memory.
\item{}can modify existing code by re-arranging, flipping the bits, buffer-over-flow, or fault injection attacks. 
\item{}cannot attack protected ROM.
\end{itemize}
\par The side-channel attack, physical access, damaging device, and control flow integrity attacks are out of scope for the proposed work.
\subsection{\bf{Design Choices}} \label{assum1}
\par \emph{SRACARE} system incorporates the following design choices to satisfy the security properties discussed in subsection~\S\ref{securityprop}.\\
 \label{features}
{\includegraphics[width=.3cm,height=.3cm]{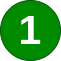}} \label{SS} {\bf{Secure Storage:}} \emph{SRACARE} uses separate ROM for storing the device information such vendor ID, Unique Device Identification (UUID), firmware revision, the symmetric cryptographic shared key ($K$) and extends the layout of the secure ROM to store a trusted recovery image.\\
{\includegraphics[width=.3cm,height=.3cm]{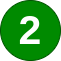}} {\bf{Frame Data Structure and Internal Communication:}} The conventional secure boot and RA systems uses an internal system bus for communication between the trusted and untrusted hardware-software modules. Attack \cite{Samebus:2017} has demonstrated that it can boot the device through malicious hardware connected to the same interconnect bus. Therefore, the proposed framework uses a dedicated SPI bus for communication between the ROM, flash, and prototype \emph{CARE} module. Moreover, the conventional system computes the digest over the entire firmware image and transfers the results via a universal interconnect bus. Since \emph{SRACARE} uses the SPI bus, if the SPI tool's hardware or software gets corrupted, it can send occasionally corrupted bits or wrong data. Therefore, the proposed design divides the flash image into 1~KB chunks. Section~\S\ref{working} covers the details of the frame data structure. This design choice is used for Proof Of Concept (POC) implementation only, and the user can parameterize frame size to optimize the performance.\\
{\includegraphics[width=.3cm,height=.3cm]{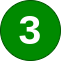}} \label{CAe}{\bf{Code Integrity \& Authentication (CA) Unit:}} Most modern secure boot architecture uses two crypto-cores - one for the digest computation and others for code signing. After performing speed, resource, and cost evaluation of various crypto-cores such as AES, RSA, ECDSA, SHA3, SHA256, SHA384, and HMAC-SHA256, the design choice of using single hardware implementation of the crypto-core HMAC-SHA256 is made for digest computation and code authentication. This hardware reuse makes the proposed solution lightweight and suitable for the targeted small embedded and IoT devices.\\
{\includegraphics[width=.3cm,height=.3cm]{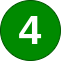}}\label{RES} {\bf{Resilience Engine (RE):}} RE provides the ability to recover the device from memory modification attacks by reflashing the affected flash memory region using onboard recovery (backup) code, It also applies the access control policies to protect the device from future attacks.\\
{\includegraphics[width=.3cm,height=.3cm]{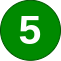}} {\bf{Trusted Execution:}} \emph{SRACARE} provides trusted execution environment by isolating the TCB and processor, and using a dedicated SPI bus for internal communication.  \\
{\includegraphics[width=.3cm,height=.3cm]{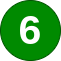}{\bf{~Secure Communication Protocol}:}} \emph{SRACARE} uses the proposed lightweight, secure communication protocol to satisfy the secure RA properties A1 and A2 from subsection~\S\ref{seccom} by leveraging existing hardware resources and following steps {1} to {12} from Fig~2. It also demonstrates a custom nonce generation technique for small embedded and IoT devices.
\vspace{-1em}
\section{\emph{SRACARE} Framework}
\subsection{\bf{System Design}} \label{SRACAREO}
\par
\begin{figure}[h]
\begin{center}
\includegraphics[width=3.4in]{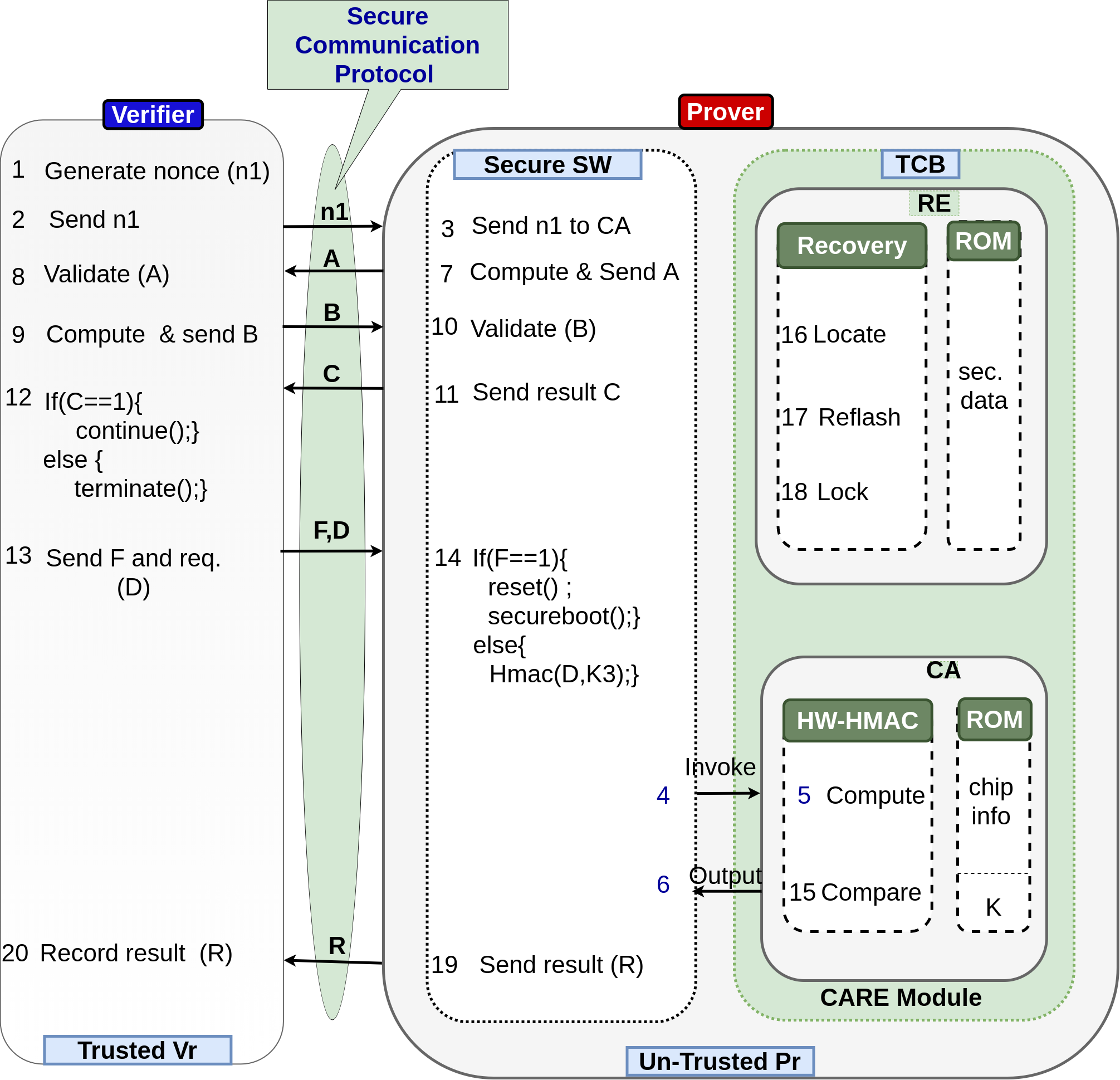}
\end{center}
\caption{Highlights the system design and key contributions of \emph{SRACARE}: 1) Novel lightweight, secure authenticated communication protocol (steps {1} to {12}), and 2) Secure boot with \emph{CARE} and remote attestation architecture for the P\textsubscript{r} device (steps {13} to {20})).
	\vspace{-1em}
\label{fig:main}}                          
\end{figure}

 Fig~\ref{fig:main} shows the top-level design overview of the \emph{SRACARE} framework, highlighted are two main security-enhancing techniques proposed by this work: 1) Lightweight, secure communication protocol and 2) Secure boot with \emph{CARE} and RA architecture for the P\textsubscript{r} device. The notations and definitions used for the communication are listed in Table~{\ref{reftab}}. \emph{SRACARE} establishes lightweight, secure communication protocol between the trusted V\textsubscript{r} and un-trusted P\textsubscript{r} by following steps {1} to {12} from Fig~\ref{fig:main}, and the detailed working is explained in subsection~\S\ref{SCP}. 
\vspace{-1.5em}
\begin{table}[H]\caption{Notations and description \label{reftab}}
\begin{tabular}{ |p{2cm}|p{5.6cm}|  } 
\hline
Notation  & Description \\
\hline       
\hline
n1   &V\textsubscript{r}'s nonce for freshness \\
n2   &P\textsubscript{r}'s nonce for freshness \\
     & n2 = Hmac(K, T) \\
     & T = hash(CHIP INFO.) $\oplus$ n1 \\
K &Symmetric key for HMAC \\
Hmac(K, m) &H(($K'\oplus 0$x$5C5C$) $||$ H(($K'\oplus 0$x$3636$) $||$ m)) \\
A &A = Hmac(K, n1) $>>$ n2  \\
B &B = Hmac(K\textsubscript{1}, n2)   \\
C &C is a true or false result of the validation of B. \\
D &D consists of parameters S\textsubscript{addr} and L as payload for attestation \\
F &Reset Flag  \\
S\textsubscript{addr} &Start address of flash memory for hashing \\
L &Lenth of the memory region to be hashed \\
R &Final Result  \\
$K'$ &$\left\{
                \begin{array}{ll}
                  H(K) \;\;K\;is\;larger\; than \; the \;block \;size  \\
                  K \;\;\;\;\;\;\; otherwise\\
                \end{array}
                \right.$ \\
m &Memory region to be attested, derived from S\textsubscript{addr}, L \\ 
H &Cryptographic hash function \\    
$K'$ &Key derived from the secret key K   \\
K\textsubscript{1} &K\textsubscript{1}= (Hmac(K, n1) $\oplus$  n1 $\oplus$ n2))   \\
$||$ &Denotes concatenation \\    
$\oplus$ &Denotes bitwise exclusive or (XOR)  \\  
CA &Code Authentication \\    
RE &Resilience Engine \\  
RA &Remote Attestation  \\  
\hline
\end{tabular}
\end{table}
The proposed secure communication protocol has two advantages over conventional authenticated communication protocols: (1) It authenticates both end devices (the P\textsubscript{r} and V\textsubscript{r}) in the communication and provides resilience from \cite{Man1}, \cite{Replay}, and \cite{DoS} attacks. (2) It does not require additional computationally heavy system resources such as TRNG, Authenticated Encryption with Associated Data (AEAD), Elliptic Curve Digital Signature Algorithm (ECDSA) or complex Message Authentication Code (MAC) to satisfy A1 \& A2 security properties listed in section~\S\ref{securityprop}. 
\begin{figure}[h]
\begin{center}
\vspace{-1em}
\includegraphics[width=3.3in]{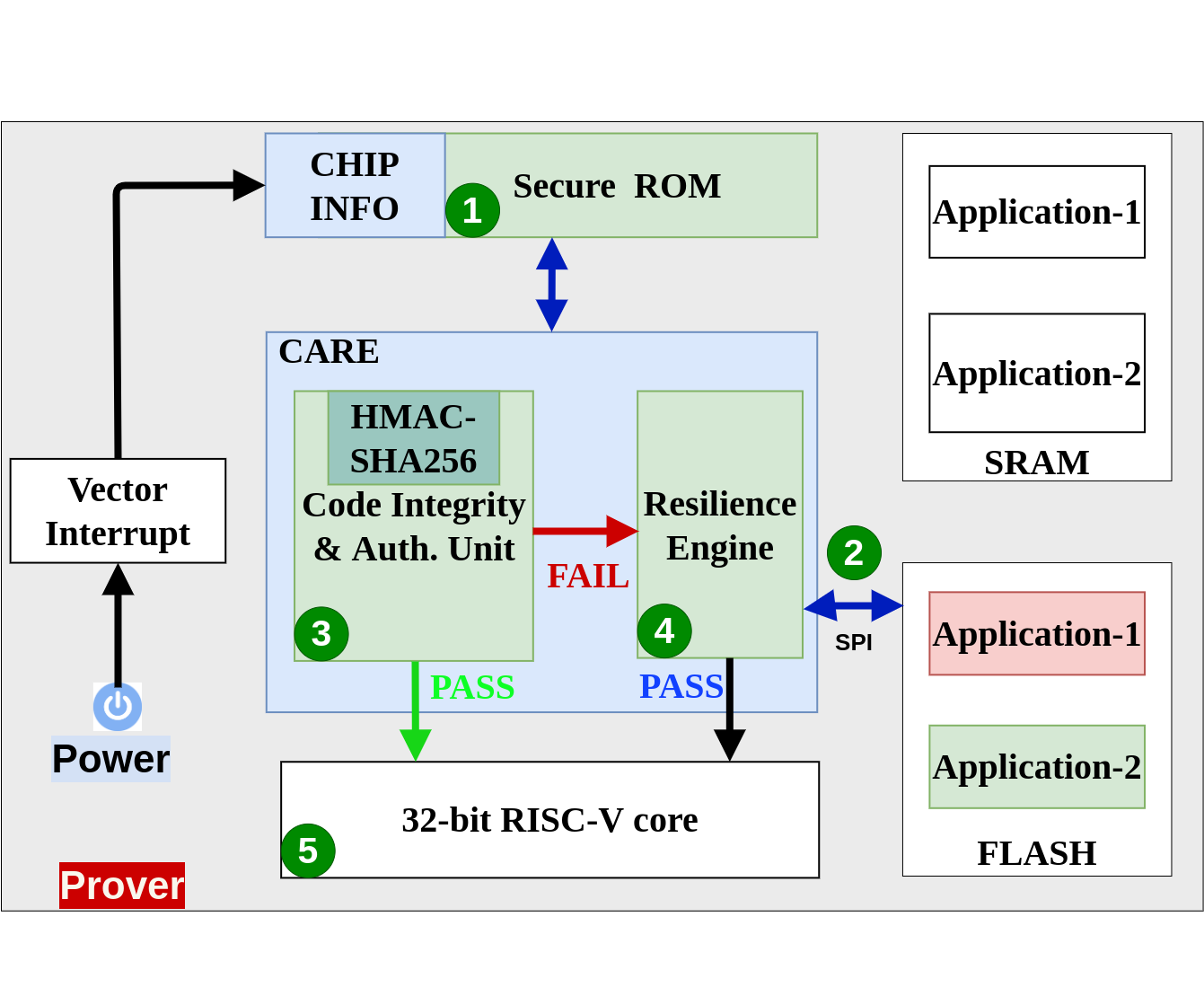}
\end{center}
\vspace{-3.5ex}
\caption{Shows the architecture design of \emph{SRACARE} based P\textsubscript{r} system, highlighted are the key design modules. The pass arrows indicate that only the known good code will be allowed to be executed on the \mbox{RISC-V} core at any given time.}
\vspace{-2.0ex}
\label{fig:care}
\end{figure} 
 To satisfy all the security properties from A3 to A8 discussed in section~\S\ref{securityprop}, \emph{SRACARE} based P\textsubscript{r} system follows design choices {\includegraphics[width=.3cm,height=.3cm]{figs/n1.png}} to {\includegraphics[width=.3cm,height=.3cm]{figs/n5}} listed in section~\S\ref{assum1}, as highlighted in Fig~3. The P\textsubscript{r} performs either the RA or secure boot with \emph{CARE} by following steps {13} to {20} from Fig~2. The detailed working of the system is covered in section~\S\ref{working}.
\subsection{\bf{System Operation}} \label{working}
\par
\begin{figure}[h]
\begin{center}
\includegraphics[width=3.3in]{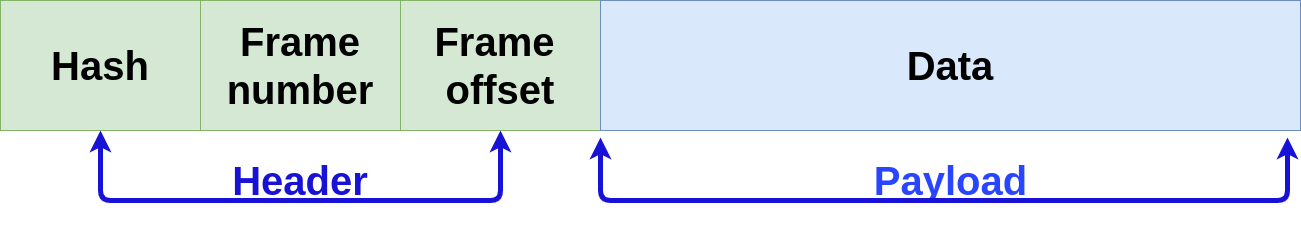}
\end{center}
\vspace{-3ex}
\caption{Represents the frame data structure. The header contains the digest of the entire frame, frame number, and flash offset location. The payload contains corresponding data for each frame.  
\vspace{-2ex}
\label{fig:frame}}
\end{figure}

 The working of \emph{SRACARE} system is divided into four main steps: 1) Secure Communication Protocol, 2) Secure Boot, 3) Resilience and Recovery, and 4) Remote Attestation. \\ 
{\bf{1) Secure Communication Protocol}:}\label{SCP} The communication starts by the V\textsubscript{r} sending nonce n1 to the P\textsubscript{r}. The P\textsubscript{r} computes Hmac(K, n1) using the crypto-core and generates n2 by performing Hmac(K, T).
\[n2 = Hmac(K, T)  \]     
\begin{equation}\label{eq23}
T = hash(CI\textsubscript{start}, 16)\;\; xor \;\;n1
\end{equation}
 Where T is calculated by taking the hash of the first 16 Bytes of the secure chip information memory and xor it with the received value of n1 (for freshness). The chip information (CHIP INFO) memory consists of the device specific information such as device serial number, firmware version, and UUID as depicted in Fig~3. Term CI\textsubscript{start} in equation~\ref{eq23} points the starting location of Chip Info (CI) memory. This novel approach gives unique n2 each time without requiring resource-heavy salt generation techniques such as TRNG. The P\textsubscript{r} generates A = (Hmac(K, n1) $>>$ n2) by appending n2 with Hmac(K,n1) and sends it to the V\textsubscript{r}. The V\textsubscript{r} validates the authenticity of the P\textsubscript{r} by recomputing Hmac(K, n1) and matching it with the received value. The V\textsubscript{r} derives the new secret key K\textsubscript{1}, computes Hmac(K\textsubscript{1}, n2), and sends the result to the P\textsubscript{r}. The P\textsubscript{r} follows the appropriate generation and validation steps to authenticate the V\textsubscript{r} and sends the result Flag C (step {11} from Fig~2) to the V\textsubscript{r}. \emph{SRACARE} closes the POC UART connection (it can be Xbee or other) between the P\textsubscript{r} and V\textsubscript{r} devices when the V\textsubscript{r} receives (C==0) (in step {12} from Fig~2), else it sends the Flag F defining the next action and associated payload to the P\textsubscript{r}. {\bf{2) Secure Boot}:} If the received Flag (F) is set (F==1), then the P\textsubscript{r} calls system reset function and performs the secure boot with \emph{CARE}. Note that steps {\color{blue}4} to {\color{blue}6} in Fig~\ref{fig:main} are represented differently to denote that those steps will be part of both RA or secure boot. However, the sequence of execution will be different. As depicted in Fig~3, the secure boot sequence starts with the system power-on. It locates and executes the First Stage Boot Loader (FSBL) code from secure ROM to initialize the SPI and flash controllers, read chip information such as - device UUID, board version, and symmetric share key, and hand off the control to the second stage boot code called the bootstrap. The bootstrapping process divides the flash image into a 1~KB frame chunks and sends it one at a time to the host via SPI bus for integrity and authenticity check. Each frame consists of the header and associated payload, as indicated in Fig~{\ref{fig:frame}}.
 \vspace{-1ex}
The header contains the digest of the entire frame, frame number, and the flash offset location. The offset location is the flash memory offset location used for the frame reflashing. The payload contains the corresponding data for each frame. This work has leveraged Hash based Message Authentication Code's (HMAC) feature for signing (authenticating) the data and SHA256 feature for integrity check for each frame to reduce hardware footprint and cost. Secure boot follows steps {\color{blue}4}-{\color{blue}5}-{15}-{16}-{17}-{18}-{\color{blue}6} from Fig~2 for each frame, and upon digest mismatch detection, the P\textsubscript{r} triggers the RE else the device will continue the normal boot process. {\bf{3) Resilience Engine}:} RE follows steps {16}-{17}-{18} from Fig~2 to locate the affected memory region, reflash the corrupted flash memory region with the known good software code from secure ROM, and lock the unauthorized access to the flash region using Physical Memory Protection (PMP) mechanism of the \mbox{RISC-V} processor. {\bf{4) Remote Attestation}:} If the received Flag (F==0) value is not set, the P\textsubscript{r} performs remote attestation based on the payload provided by the V\textsubscript{r}, which consists of the start location and the length of the information to be attested. The P\textsubscript{r} follows steps {\color{blue}4}-{\color{blue}5}-{\color{blue}6} sequence from Fig~2 to compute the digest and it sends the report to the V\textsubscript{r} (steps {19} and {20} from Fig~2).

\section{Evaluation} \label{Evaluation}
This section describes the chain-of-trust theory, resource utilization, and performance analysis for each submodule in \emph{SRACARE} framework design, and comparison with state-of-the-art solutions.  
\vspace{-.7em}
\subsection{\bf{Chain-of-Trust}}\label{chainoftrust}
The work presented in \cite{Arbaugh:1997} defines the secure boot process as a chain of many small layers of the boot codes executed in a specific sequence. It requires the boot process to follow two rules for the integrity measurement assurance: {{\bf 1)}~The integrity is checked for all the lower layers.} {{\bf 2)}~Transitions to higher layers occurs-only after integrity checks on all lower layers are completed.} The conventional secure boot systems measure the integrity of each stage (layer) of the boot code at the file level. The proposed work uses the same concept in a different context as it breaks down the entire image into 1~KB chunks (called frames) and measures code integrity and authenticity of each frame. Following equation represents the chain-of-trust for \emph{SRACARE} system: 
\vspace{-1ex}
\[ I_0{} = True \]
\vspace{-2em}
\begin{equation}
I\textsubscript{\emph{i}+1} = I_\emph{i}{}\;\;\;\; \& \;\;\;\; V_\emph{i}{}(L\textsubscript{\emph{i}+1} )
\end{equation}
$I_\emph{i}{}$ and \& are the boolean value representation of the integrity of frame \emph{i} and AND operation respectively. The verification function associated with the $\emph{i}^{th}{}$ layer is represented by $V_\emph{i}{}$. $V_\emph{i}{}$ takes the layer to verify as its only argument and returns boolean result. It performs a cryptographic hash of the frame and compares the result with stored digest value. As explained earlier, this work has divided boot flash data into 1~KB frame chunks, and hash digest is computed on each frame for verification. Therefore, the recurrence is represented by:
\begin{equation}
   I\textsubscript{\emph{i}+1} =\left\{
                \begin{array}{ll}
                  I_\emph{i=0}{}  = True \;\;\;\;\;\;\;\;\;\;\;\;\;\;\;\; for \;\; \emph{i} = 0\\
                  
                  I_\emph{i}{} \;\; \& \;\; V_\emph{i}{} (L\textsubscript{\emph{i}+1}) \;\;\;for \;\; \emph{i} = 1,2,3 ...n\\
                  
                \end{array}
              \right.
 \end{equation}
Here, $I_0{}$ is the trusted boot ROM code, and the integrity of ROM is taken as boolean true. $n$ represents the number of frames of the flash image. The following equation~\ref{eq3} calculates the estimated increase in boot time (T\textsubscript{$\Delta$}) for a secure boot with \emph{SRACARE}.  
\begin{equation}\label{eq3}
T\textsubscript{$\Delta$} = t\textsubscript{fm\_0}(V_0{}(L_1{}))\;\; + t\textsubscript{fm}(\sum_{\emph{i}=1}^{n}V_\emph{i}{}(L_{\emph{i}+1}{}))
\end{equation}
\vspace{-1em}
\par Where t\textsubscript{fm\_0} is the execution time for the first frame, and t\textsubscript{fm} is the execution time for all remaining frames. The verification time includes the time required to compute and verify the message digest. By design, \emph{SRACARE} first matches the frame number of the received frame and clears the flash region to reflash it with a trusted code. Therefore, the first frame  processing requires more time (t\textsubscript{fm\_0}) than the remaining frames.
\subsection{\bf{Code Authentication (CA) Unit}}\label{PA}
\par The key component of the Code Authentication (CA) unit is crypto-core (HMAC-SHA256). To estimate the initial performance, power, and resource utilization, our evaluation setup uses both hardware and software implementation of
\vspace{-.8em}
\begin{table}[h]
\caption{Performance Analysis of CA on FPGA.}\vspace{-.7em}
	\scalebox{1.4}[1.2]{
	\begin{tabular}{@{}lcc@{}}
	\toprule
	\multicolumn{3}{c}{Performance Analysis of crypto-core on FPGA}  \\ \midrule
	\multicolumn{1}{l|}{Parameters}              & \multicolumn{1}{c|}{Software} & Hardware \\ \midrule
	\multicolumn{1}{l|}{Cycles (c)}         & \multicolumn{1}{c|}{\bf{47033}}   &  {\bf{2926}}   \\
	\multicolumn{1}{l|}{Frequency (f)(MHz)}            & \multicolumn{1}{c|}{100}   & 100    \\
	\multicolumn{1}{l|}{Block (b)} & \multicolumn{1}{c|}{256}      & 256    \\
	\multicolumn{1}{l|}{Throughput T(Mbps)}     & \multicolumn{1}{c|}{.54}    & 8.74   \\
	\multicolumn{1}{l|}{Time (usec)}           & \multicolumn{1}{c|}{470.33}       & 29.26    \\
	\multicolumn{1}{l|}{Energy Consumption (E) }                & \multicolumn{1}{c|}{197.06}       & 12.25    \\
	\multicolumn{1}{l|}{Energy Efficiency}      & \multicolumn{1}{c|}{\bf{92.68}}    & {\bf{0.358}}    \\ \bottomrule
	\vspace{-2.3em}
	\end{tabular}
	\vspace{-3em}
	}\label{tab:hwswperf}
\end{table}  
the crypto-core running on the baremetal \mbox{RISC-V} processor to compute the digest of same data size of 256 Bytes. Table~\ref{tab:hwswperf} illustrates a performance increase of 16x and 92${\%}$ less power utilization while using the hardware HMAC-SHA256. Also, software implementation requires 3.6~KB additional secure storage and assumes the \mbox{RISC-V} core to be part of TCB. Including the processor in TCB is not a preferred design choice due to the processor related vulnerabilities. Therefore, a hardware-based crypto-core is selected for the CA unit.
\vspace{-.8em}
\subsection{\bf{Resilience Engine (RE)} }\label{swres}\vspace{-.4em}
\par The Resilience Engine (RE) is implemented in software for the POC work. For the test application of 5.6~KB, it requires {\bf{61}} additional lines of code (C language) in the secure boot code base and approximately 5~KB of additional secure ROM to store the golden recovery image data. The Resilience Engine (RE) requires 968~bytes of recovery data for every 1~KB of the flash image. The secure ROM size increases exponentially with the size of the application code used for recovery. To limit the size of the recovery data storage, the application developer can select the necessary code module for the recovery process or use a suitable compression technique, to bring the device to the bare minimum working state. Although, this feature is not implemented in the proposed work as the test application uses only 5~KB of additional ROM for recovery. 
\vspace{-.5em}
\subsection{\bf{System Performance}}\vspace{-.5em}
\par For \emph{SRACARE} system {\bf{timing analysis}}, a test application of 5.6~KB is divided into six 1~KB chunks and the total time for the system boot-up with and without the proposed framework is calculated and presented in Table~\ref{tab:timeanalysis}. The framework uses equation~(\ref{eq3}) to calculate the total execution time $T$ and equation~(\ref{eq4}) for time difference D\textsubscript{$\Delta$}. As seen from Table~\ref{tab:timeanalysis}, the time difference in the total time D\textsubscript{$\Delta$} = .529 milliseconds indicates that proposed \emph{SRACARE} architecture requires 8 percent extra boot-time, 5~KB extended ROM and increases bootstrap code by 61 lines, which are insignificant in comparison to the level of security, onboard recovery, and resilience it provides.
\vspace{-1em}
\begin{table}[H]
	\caption{Timing analysis for Bootstrap.}
	\scalebox{1.1}[1.1]{
	\begin{tabular}{@{}lcc@{}}
	\toprule
	\multicolumn{3}{c}{Timing Analysis of secure bootstrap on FPGA}  \\ \midrule
	\multicolumn{1}{l|}{Parameters}              & \multicolumn{1}{c|}{Without} & With \\ \midrule
	\multicolumn{1}{l|}{Cycles req. for the first frame (c)}         & \multicolumn{1}{c|}{553611}   &  576083   \\
	\multicolumn{1}{l|}{Cycles (rest of frames)(c)}       & \multicolumn{1}{c|}{103330}   & 133790    \\
	\multicolumn{1}{l|}{Total Cycles}            & \multicolumn{1}{c|}{\bf{656941} }   & \bf{709873}   \\
	\multicolumn{1}{l|}{Frequency (f)(MHz)}            & \multicolumn{1}{c|}{100}   & 100    \\
	\multicolumn{1}{l|}{Time (T) (usec)}           & \multicolumn{1}{c|}{\bf{6569.41}}       & \bf{7098.73}    \\ \midrule
	Time difference D\textsubscript{$\Delta$} = {\bf{529.32}} usec \\
	which is 8\% more than without secure boot\\ \bottomrule
	\vspace{-3em} 
	\end{tabular}
	}
	\label{tab:timeanalysis}
\end{table} 
\vspace{-1em}
\begin{equation}\label{eq4}
D\textsubscript{$\Delta$} = Time (with \emph{SRACARE}) - Time (without \emph{SRACARE}) 
\end{equation}
\vspace{-1.2em}
\subsection{\bf{Resource Utilization}}
Inspired from the googles opentitan \cite{Ope:2019}, the prototype implementation uses the Ibex \mbox{RISC-V} core for the RTL design.
\vspace{-1em}
\subsubsection{\bf{Hardware Resource Utilization on FPGA}}
\begin{table}[h]
	\caption{Hardware Resource Utilization.}\label{qunt1}
	\vspace{-.6em}
	\scalebox{1.15}[1.1]{
	\begin{tabular}{@{}lcccc@{}}
	\toprule
	\multicolumn{5}{c}{FPGA Hardware Resource Utilization Report}  \\ \midrule
	\multicolumn{1}{l|}{Parameters}              & \multicolumn{1}{c|}{\emph{SRACARE}} & \multicolumn{1}{c|}{HMAC} & \multicolumn{1}{c|}{Ibex} & {${\%Util.}$} \\ \midrule
	\multicolumn{1}{l|}{Slice LUTs}         & \multicolumn{1}{c|}{\bf{24249}}  & \multicolumn{1}{c|}{\bf{2807}} & \multicolumn{1}{c|}{3581} &  18   \\
	\multicolumn{1}{l|}{LUT as logic}         & \multicolumn{1}{c|}{24081}  & \multicolumn{1}{c|}{2807} & \multicolumn{1}{c|}{3581} &  18   \\
	\multicolumn{1}{l|}{LUT as DRAM}         & \multicolumn{1}{c|}{168}  & \multicolumn{1}{c|}{0} & \multicolumn{1}{c|}{8} &  1   \\ \midrule
	\multicolumn{1}{l|}{Slice Registers}         & \multicolumn{1}{c|}{19586}  & \multicolumn{1}{c|}{2312} & \multicolumn{1}{c|}{2559} &  7   \\
	\multicolumn{1}{l|}{Register as FF}         & \multicolumn{1}{c|}{19581}  & \multicolumn{1}{c|}{2312} & \multicolumn{1}{c|}{2559} &  7   \\
	\multicolumn{1}{l|}{Register as Latch}         & \multicolumn{1}{c|}{5}  & \multicolumn{1}{c|}{0} & \multicolumn{1}{c|}{0} &  $<$1   \\ \midrule
	\multicolumn{1}{l|}{F7 Muxes}         & \multicolumn{1}{c|}{1407}  & \multicolumn{1}{c|}{71} & \multicolumn{1}{c|}{265} &  2   \\
	\multicolumn{1}{l|}{F8 Muxes}         & \multicolumn{1}{c|}{196}  & \multicolumn{1}{c|}{29} & \multicolumn{1}{c|}{0} &  1   \\ \bottomrule
	\end{tabular}
	}
	\vspace{-1mm}
\end{table}
Table~\ref{qunt1} depicts resource utilization of (crypto-core) HMAC-SHA256, Ibex core, complete \emph{SRACARE} design, and percentage utilization of available hardware resource on Artix~7 FPGA. The crypto-core uses 3x less area and operates at 2x faster speed compared to other implementation from \cite{Juliato2011FPGAIO}.  
\subsubsection{\bf{Software Resource Utilization}} \label{swres}
The POC work has implemented the RE submodule and secure communication protocol in software. It requires 61 additional lines of (C language) code for RE (which includes cycle calculation and analysis code for Ibex) and 5~KB of extended ROM storage.It requires 108 additional lines of code (C language) on the P\textsubscript{r} side for secure RA (UART) protocol and novel n2 generation logic implementation. \emph{SRACARE} increases the total software code base by 10\%.
\subsection{\bf{Comparison with the State-of-the-art Solutions}}
\par Since \mbox{RISC-V} is a relatively new architecture, authors of this work did not found any {\bf{secure boot with resilience or recovery}} implementation for state-of-the-art comparison. Therefore, this work compares the proposed \emph{SRACARE} framework with other state-of-the-art secure boot and RA architectures targeting \mbox{RISC-V} based embedded and IoT systems. The comparison focuses on both qualitative and quantitative analysis. Table~\ref{qunt11} illustrates the qualitative comparison. 
\subsubsection{\bf{Qualitative Comparison}}
\vspace{-1em}
\begin{table}[H]
	\caption{Qualitative comparison between secure boot/RA techniques targeting lightweight embedded devices.}\label{qunt11}
	\scalebox{.95}[0.9]{
	\begin{tabular}{@{}lccccc@{}}
	\toprule
	\multicolumn{1}{l|}{Parameters}        & \multicolumn{1}{c|}{\emph{SRACARE}} & \multicolumn{1}{c|}{Healed} & \multicolumn{1}{c|}{Ref.\cite{Secerase:2010}} & \multicolumn{1}{c|}{Ref.\cite{Haj:2019}} & \multicolumn{1}{c}{Sanctum} \\ \midrule
	\multicolumn{1}{l|}{Design Type}         & \multicolumn{1}{c|}{Hybrid}  & \multicolumn{1}{c|}{SW} & \multicolumn{1}{c|}{Hybrid} & \multicolumn{1}{c|}{HW} & \multicolumn{1}{c}{Hybrid}   \\
	\multicolumn{1}{l|}{Secure RA}         & \multicolumn{1}{c|}{yes}  & \multicolumn{1}{c|}{no} & \multicolumn{1}{c|}{no} & \multicolumn{1}{c|}{no} & \multicolumn{1}{c}{yes}   \\
	\multicolumn{1}{l|}{Detection}         & \multicolumn{1}{c|}{yes}  & \multicolumn{1}{c|}{yes} & \multicolumn{1}{c|}{yes} &  \multicolumn{1}{c|}{yes} & \multicolumn{1}{c}{yes}   \\
	\multicolumn{1}{l|}{Protection}         & \multicolumn{1}{c|}{yes}  & \multicolumn{1}{c|}{no} & \multicolumn{1}{c|}{yes} &  \multicolumn{1}{c|}{yes} & \multicolumn{1}{c}{yes}  \\ 
	\multicolumn{1}{l|}{Recovery}         & \multicolumn{1}{c|}{yes}  & \multicolumn{1}{c|}{yes} & \multicolumn{1}{c|}{yes} &  \multicolumn{1}{c|}{partial} & \multicolumn{1}{c}{no}   \\
	\multicolumn{1}{l|}{Secure boot}         & \multicolumn{1}{c|}{yes}  & \multicolumn{1}{c|}{no} & \multicolumn{1}{c|}{no} &  \multicolumn{1}{c|}{yes} & \multicolumn{1}{c}{yes}   \\  \bottomrule
	\end{tabular}
	}
   \vspace{-1mm}
\end{table}
\par Table~\ref{qunt11} shows that all techniques provide memory attack detection methods, and only \emph{SRACARE}, Healed, and \cite{Secerase:2010} provides resilience and recovery techniques. Sanctum \cite{Sanctum:2018} uses hybrid secure enclaves for code execution and RA. \cite{Haj:2019} uses hardware-based memory attack detection and protection method. Healed \cite{Healed:2019} implements RA with recovery using Markle Hash Tree (MHT). It assumes that at least one device on the network is untampered, and the code of that device can be used for recovery. \cite{Secerase:2010} provides a recovery method using a trusted ROM to store transmitter and receiver code for associating with a recovery server. Both \cite{Sanctum:2018} and \cite{Haj:2019} uses Rocket chip \mbox{RISC-V} core with OS support and complex designing. However, since \cite{Haj:2019} provides partial recovery support it is close candidate for \emph{SRACARE} comparison. 
\subsubsection{\bf{Quantitative Comparison}}\label{quantify}
\vspace{-1em}
\begin{table}[h]\caption{Quantitative comparison with state-of-the-art SECURE BOOT / REMORT ATTESTATION (RA) technique for \mbox{RISC-V}.}\label{care1} 
	\scalebox{1.05}[1.0]{
	\begin{tabular}{@{}lcccc@{}}
	\toprule
	\multicolumn{1}{l|}{Parameters}              & \multicolumn{1}{c|}{\emph{SRACARE}}  & \multicolumn{1}{c|}{HMAC} & \multicolumn{1}{c}{CAU's ECDSA only}  \\ \midrule
	\multicolumn{1}{l|}{Slice LUTs}         & \multicolumn{1}{c|}{24249}  & \multicolumn{1}{c|}{2807}  & \multicolumn{1}{c}{27170}  \\
	\multicolumn{1}{l|}{LUT as logic}        & \multicolumn{1}{c|}{24081}  & \multicolumn{1}{c|}{2807}  & \multicolumn{1}{c}{26450}  \\
	\multicolumn{1}{l|}{LUT as DRAM}         &  \multicolumn{1}{c|}{168}  & \multicolumn{1}{c|}{0}   & \multicolumn{1}{c}{720}   \\ \midrule
	\multicolumn{1}{l|}{Slice Registers}         & \multicolumn{1}{c|}{19586}  & \multicolumn{1}{c|}{2312}  & \multicolumn{1}{c}{6722}   \\
	\multicolumn{1}{l|}{Register as FF}         & \multicolumn{1}{c|}{19581}  & \multicolumn{1}{c|}{2312}  & \multicolumn{1}{c}{6722}    \\
	\multicolumn{1}{l|}{Register as Latch}         & \multicolumn{1}{c|}{5}  & \multicolumn{1}{c|}{0}   & \multicolumn{1}{c}{0}   \\ \midrule
	\multicolumn{1}{l|}{Multiplexer}    & \multicolumn{1}{c|}{}   & \multicolumn{1}{c|}{}   & \multicolumn{1}{c}{} \\
	\multicolumn{1}{l|}{F7 Muxes}         & \multicolumn{1}{c|}{1407}  & \multicolumn{1}{c|}{71} & \multicolumn{1}{c}{684}   \\
	\multicolumn{1}{l|}{F8 Muxes}         &  \multicolumn{1}{c|}{196}  & \multicolumn{1}{c|}{29}  & \multicolumn{1}{c}{0}   \\ \bottomrule
	\end{tabular}
	}	
\end{table}
\par Table~\ref{care1} shows that \cite{Haj:2019} requires 27170 slice LUTs for hardware crypto-core module (ECDSA) implementation, which is {\bf{14x}} larger than \emph{SRACARE}'s crypto-core hardware requirement. Furthermore, \cite{Haj:2019} requires two 64 bit the \mbox{RISC-V} cores for Trusted Execution Environment (TEE) implementation, hardware SHA3 for hashing, and configurable LFSR-based Physical Unclonable Function (CoLPUF) for key generation, ECDSA core for asymmetric signing, boot sequencer, and key management unit. These make \cite{Haj:2019} a resource-heavy solution and unsuitable for small embedded and IoT devices. The comparison of asymmetric and symmetric cryptographic hardware requirements provide an initial estimation of the overall hardware overhead requirements.
\vspace{-.5em}
\section{Conclusion}
\vspace{-.5em}
\emph{SRACARE} demonstrates the POC framework that performs the run-time and boot-time integrity and authenticity checks, secure boot with onboard recovery, and remote attestation on the \mbox{RISC-V} based small embedded and IoT devices. It implements the novel lightweight,  secure authenticated communication protocol. It provides the tools for the detection, protection and recovery from malicious code modification attacks. The experimental results show good protection against malicious code modification attacks, with only 8\% execution time overhead and a tiny increase in resource footprint. 
\vspace{-.8em}

\bibliographystyle{IEEEtran}

\bibliography{references}

\end{document}